\documentclass[aip,citeautoscript,floatfix,reprint,amsmath,amssymb,superscriptaddress]{revtex4-1}%

\usepackage{graphicx}
\usepackage{dcolumn}
\usepackage{bm}
\usepackage{times}

\usepackage{titlesec}
\titleformat{\section}
  {\normalfont\sffamily\large\bfseries}
  {\thesection}{1em}{}

\begin{document}

\title{Physical ageing of spreading droplets in a viscous ambient phase}

\author{Bibin M. Jose}
\author{Dhiraj Nandyala}
\author{Thomas Cubaud}
\affiliation {Department of Mechanical Engineering, Stony Brook University, Stony Brook, NY 11794, USA}
\author{Carlos E. Colosqui}
\email[]{carlos.colosqui@stonybrook.edu}
\affiliation{Department of Mechanical Engineering, Stony Brook University, Stony Brook, NY 11794, USA}
\affiliation{Department of Applied Mathematics \& Statistics, Stony Brook University, Stony Brook, NY 11794, USA}

\begin{abstract}
Nanoscale topographic features of solid surfaces can induce complex metastable behavior in colloidal and multiphase systems.
Recent studies on single microparticle adsorption at liquid interfaces have reported a crossover from fast capillary driven dynamics to extremely slow kinetic regimes that can require up to several hours or days to attain thermodynamic equilibrium.  
The observed kinetic regime resembling physical ageing in glassy materials has been attributed to unobserved surface features with dimensions on the order of a few nanometers. 
In this work, we study the spontaneous spreading of water droplets immersed in oil and report an unexpectedly slow kinetic regime not described by previous spreading models.
We can quantitatively describe the observed regime crossover and spreading rate in the late kinetic regime with an analytical model considering the presence of periodic metastable states induced by nanoscale topographic features (characteristic area $\sim$4 nm$^2$, height $\sim$1 nm) observed via atomic force microscopy.
The analytical model proposed in this work reveals that certain combinations of droplet volume and nanoscale topographic parameters can significantly hinder or promote wetting processes such as spreading, wicking, and imbibition.
\end{abstract}

\maketitle

%
Classical continuum descriptions consider liquid-fluid and liquid-solid interfaces as sharp, smooth, and homogeneous surfaces, which neglects the diffuse nature of the interfacial region, the presence of nanoscale heterogeneities of physical or chemical origin, and thermal fluctuations.\cite{gibbs1906,rowlinson,degennes1985,brochard1991} 
Despite remarkable successes in rationalizing the dynamics of wetting and related interfacial phenomena, classical continuum-based models are inadequate to describe the near-equilibrium behavior of diverse colloidal and multiphase systems where the interplay between thermal motion and nanoscale interfacial structure plays a dominant role.
For example, single microparticles adsorbed at liquid-liquid interfaces have exhibited crossovers from initially fast dynamics, driven by capillary forces, to a much slower ``kinetic'' relaxation that can be nearly logarithmic in time, as first discovered by Kaz {\it et al.}\cite{kaz2012} and subsequently studied by other groups.\cite{colosqui2013,rahmani2016,coertjens2017,zanini2017,keal2017} 
Similar near-equilibrium behavior has been also observed in the imbibition/drainage of water/oil in microscale capillaries with nanoscale surface roughness.\cite{prf2016}
The observed near-equilibrium phenomena, resembling physical ageing in dense colloidal systems that exhibit jamming transitions,\cite{struik1977,fluerasu2007,negi2009,ovarlez2010} have been attributed to random thermally activated transitions between multiple metastable states induced by numerous nanoscale ``defects'' on the solid surface.

%

\begin{figure*}
\center
\includegraphics[angle=0,width=1\linewidth]{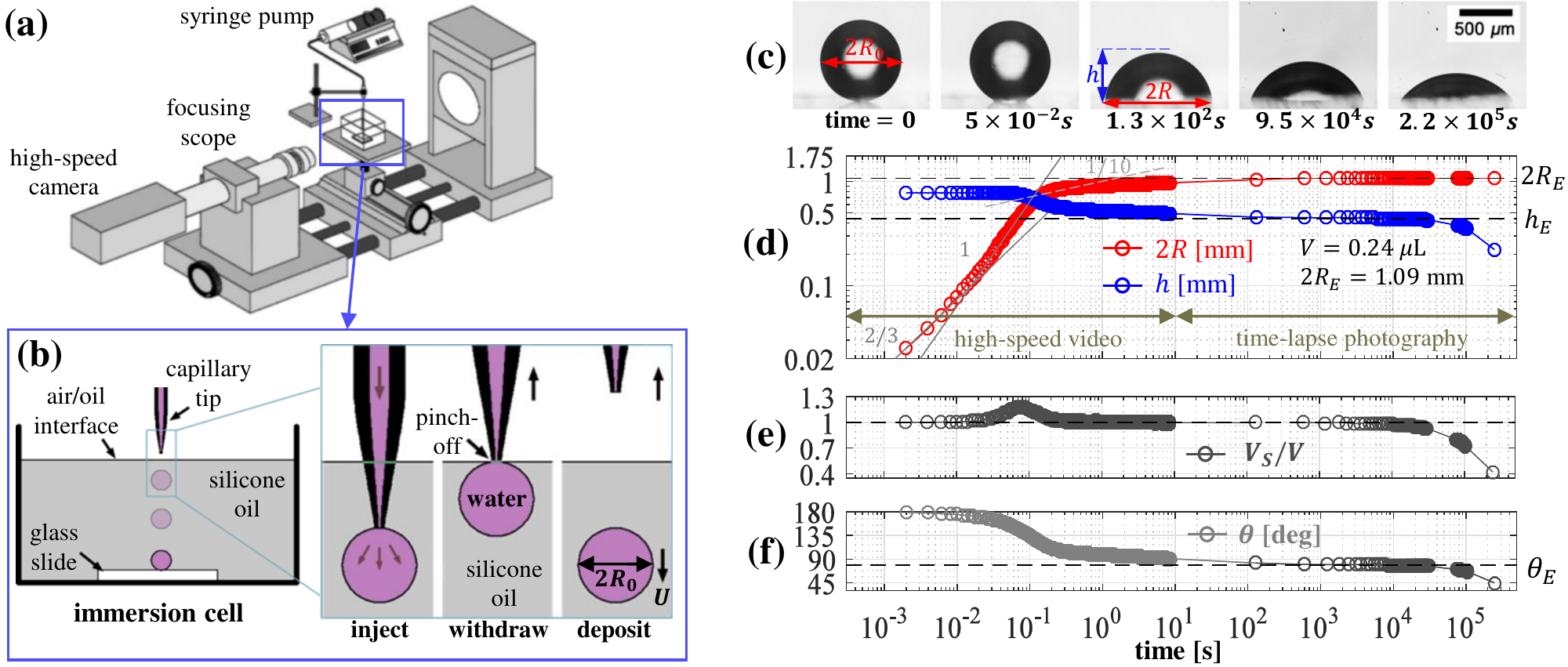}
\vskip -5pt
\caption{Experimental procedure and analysis.
(a) Experimental apparatus to study spreading of water droplets in 100-cSt silicone (PDMS) oil.
(b) Sequence of steps leading to the gentle deposition of a water droplet on a borosilicate slide. 
(c-d) Observed time evolution of contact radius $R$ and height $h$ obtained from digital images recorded at 32 fps during 10~s ($R_0=0.4$~mm, $V=0.27~\mu$L). 
Power-law fits are included for comparison.
(e-f) Volume $V_S=\pi h (R^2/2+h^2/6)$ and contact angle $\theta$ corresponding to a spherical cap.
}
\label{fig:1}
\end{figure*}

Considering the physical mechanisms involved, similar near-equilibrium behavior and physical ageing phenomenon observed for single colloidal particles and microcapillaries should be expected in the classical problem of droplet spreading, which is relevant to technical applications ranging from water treatment and oil recovery to biochemical assay and 3D printing.
Although the dynamics of spontaneous droplet spreading has been extensively investigated,
\cite{tanner1979,chen1988experiments,chen1992edge,leger1992liquid,zosel1993studies,de1997contact,ruijter1999,mchale2004topography,davidovitch2005,mchale2009} to the best of our knowledge no previous study has reported a crossover from capillary driven dynamics and power-law spreading to slow logarithmic relaxation and physical ageing behavior and analogous to that observed for the adsorption of single colloidal particles.\cite{kaz2012,colosqui2013,rahmani2016,keal2017}  
Spreading studies have established that the droplet contact radius $R$ in different dynamic regimes follows a power law $R(t) \propto t^\alpha$ with exponents $\alpha$ varying from 0.1 to 1.
Spreading regimes where $R(t)\propto t^{0.5}$ have been observed at short initial times where inertial effects are significant, for droplets of liquids with low viscosity ($\lesssim$ 10 mPa$\cdot$s) immersed in a gas/vapor phase.\cite{biance2004first,bird2008} 
For droplets of liquids with higher viscosity ($\gtrsim$ 100 mPa$\cdot$s), viscous damping can dominate even at short initial times due to the formation of a very small meniscus below the droplet and nontrivial power laws with variable exponents $\alpha \simeq$~1 to 0.1 have been observed.\cite{paulsen2011,eddi2013}
On the other end, the late spreading dynamics of small viscous droplets is commonly overdamped and characterized by Tanner's law $R(t) \propto t^{1/10}$.\cite{tanner1979}
It is worth noticing that a power-law $R(t) \propto t^{1/10}$ is not to be expected for any of the conditions studied in this work since Tanner's law is derived using thin film lubrication theory and is strictly valid for small contact angles (i.e., below 40$^\circ$) and negligible viscosity of the ambient phase.    
While numerous previous works have studied droplet spreading within a much less viscous ambient phase, a significantly smaller number of experimental studies have investigated the dynamics of liquid droplets immersed in a much more viscous ambient phase.\cite{mitra2016,jose2017} 
For the latter case, the droplet contact radius has been found to still follow power laws $R(t) \propto t^{\alpha}$ with exponents $\alpha$ that present a nontrivial dependence on the viscosity of each phase.\cite{jose2017}
It has been shown, in particular, that the external phase viscosity can strongly influence the droplet spreading dynamics.\cite{jose2017} 
%
%
The different power laws observed in the dynamics of droplet spreading reflect the dominance of different physical mechanisms for energy dissipation and associated damping forces. 
Such damping forces have been commonly estimated using classical models for contact line dynamics such as the Voinov-Cox model\cite{voinov1976,cox1986}, which considers viscous hydrodynamic friction, and the so-called molecular kinetic theory (MKT)\cite{blake1969}, which considers that irreversible work is performed in the adsorption and desorption of molecules at a moving contact line.
However, available models for contact line dynamics cannot account for the dynamic-to-kinetic regime crossover and extremely slow near-equilibrium relaxation reported for systems involving the displacement of liquid-liquid interfaces on a solid, as in the case of micro/nanoparticle adsorption and micro/nanocapillary imbibition.\cite{kaz2012,colosqui2013,rahmani2016,keal2017,prf2016}

\begin{figure*}
\center
\includegraphics[angle=0,width=1.\linewidth]{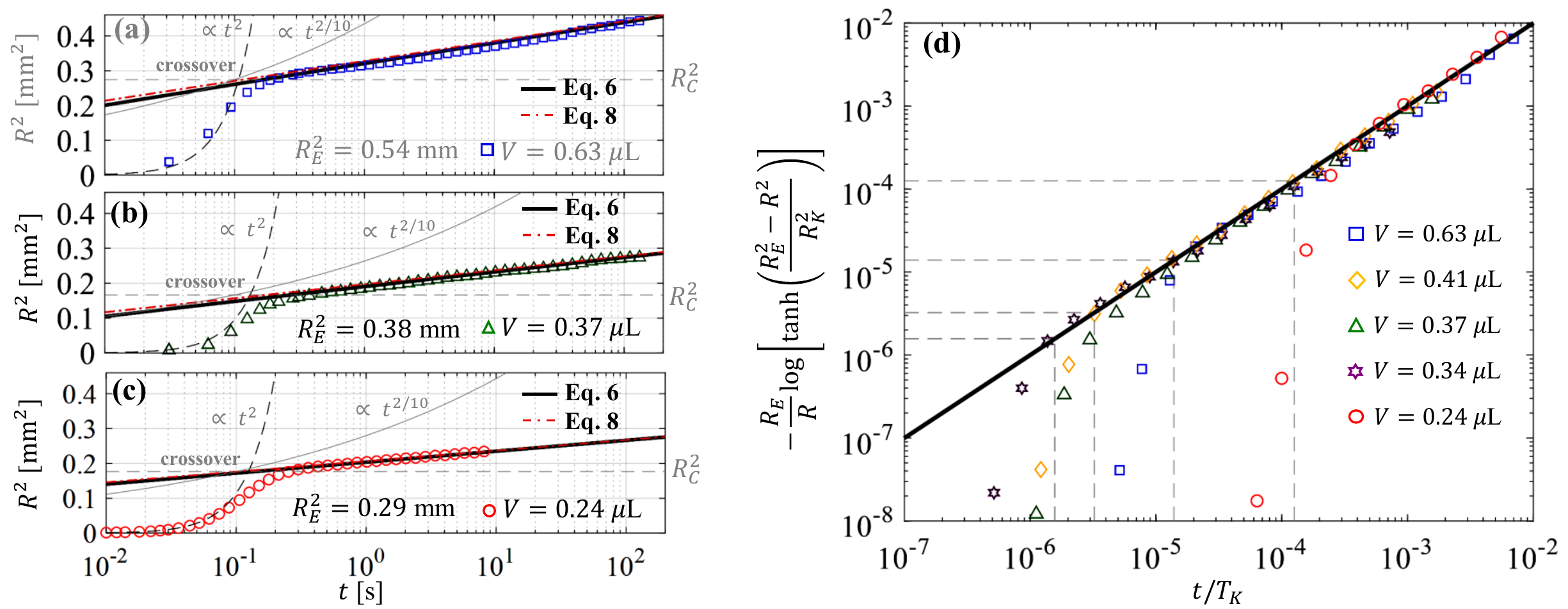}
\vskip -5pt
\caption{Experimental results and analytical predictions.
(a-c) Square contact radius $R^2$ vs. time $t$ for droplets of volume $V=$~0.24, 0.37, \& 0.63 $\mu$L.
Markers correspond to experimental results. 
The crossover radius $R_C$ is predicted by equation (\ref{eq:crossover}) for $\alpha=0.55$. 
$R_E$ is the expected equilibrium contact radius for each case. 
Power-law fits $R \propto t$ and $R \propto t^{0.1}$ are included for comparison.
(d) Experimental data for droplets with different volumes (markers) scaled according to equation (\ref{eq:area}). Time is normalized by the kinetic time defined in equation (\ref{eq:Tk}). Dashed lines are predictions from equation (\ref{eq:crossover}) for the crossover to kinetic spreading. 
Analytical fits via equations (\ref{eq:area})--(\ref{eq:crossover}) employ the parameters reported in Table 1.   
}
\label{fig:2}
\end{figure*}

\noindent {\bf \large Spreading in a viscous ambient phase}\\
\noindent This work investigates the near-equilibrium behavior of millimeter-sized water droplets immersed in 100-cSt silicone oil and spreading spontaneously on a borosilicate glass substrate, for which nearly neutral wetting conditions are attained at equilibrium. 
Our spreading experiments are performed inside a transparent immersion cell (see Fig.~\ref{fig:1}a) that is placed on the stage of an optical goniometer (Dataphysics OCA 10).
The immersion cell is filled with 100-cSt silicone oil with a dynamic viscosity $\mu_o=96$mPa$\cdot$s (mass density $\rho_o=0.96$~g/mL) and the top of the cell is open to the ambient air at room temperature ($T=20\pm 4~C^\circ$).
As illustrated in Fig.~\ref{fig:1}b, single drops of de-ionized (DI) water are injected into the oil bath through a tapered capillary tube (inner tip diameter $\sim$ 0.04 mm) that is connected to a programmable syringe pump.
A surface tension $\gamma=42.6\pm 2$~mN/m between the silicone oil and DI water at room temperature and static equilibrium contact angles $\theta_E=74.5\pm 3^\circ$ were steadily observed for about 24 hours in Wilhelmy plate measurements (see Methods section).
The injected drops have a radius $R_0\simeq 0.5$~mm, which correspond to very small Bond numbers 
$Bo=(\rho_w-\rho_o)g R_0^2/\gamma\simeq 0.002$; here $\rho_w$ is the water mass density and $g$ is the gravitational acceleration.
After detaching from the capillary, water droplets fall toward the glass slide on the cell bottom (see Fig.~\ref{fig:1}b), attaining a small terminal speed 
$U=2(\rho_w-\rho_o)g R_0^2/9\mu_o\simeq 1$~mm/s in agreement with Stokes flow predictions \cite{hadamard1911} for a sedimenting spherical droplet ($R_0=0.5$~mm and viscosity ratio 1:100).
%
%
%
The terminal speed of the drop corresponds to very small Weber numbers 
$We =\rho_w U^2 R/\gamma \sim 10^{-5}$, which indicates that inertial effects are negligible and the droplet deposition on the glass substrate can be considered as quasi-static.
The spontaneous spreading of the deposited droplets begins immediately after making contact with the glass substrate.
The time evolution of the droplet contact radius $R(t)$ and height $h(t)$ (see Fig.~\ref{fig:1}c-d) are recorded by combining high-speed video during the initial 0.002 to 130 seconds and time-lapse photography for up to 3 days in order to efficiently record the (fast) initial and (slow) late spreading regimes.
We begin our analysis of the experimental results by assessing the classical modeling assumption that the studied droplets, characterized by very low Bond numbers $Bo\ll 1$, maintain the shape of a spherical cap with constant volume $V$ during the spreading process.
The volume $V=(4/3)\pi R_0^3$ of the droplets is readily determined from their initial radius $R(t=0)=R_0$ obtained from acquired images (cf. Fig.~\ref{fig:1}c).
Experimental observations indicate that the droplet contact radius $R(t)$ and height $h(t)$ (cf. Fig.~\ref{fig:1}d-e) correspond to those of a spherical cap of nearly constant volume $V_S(t)=\pi h(R^2/2+h^2/6)\simeq V$ for times between 0.5 and $10^4$ seconds (cf. Fig.~\ref{fig:2}c).
Outside this time window deviations from spherical shape (cf. Figs.~\ref{fig:1}c-e) are observed due to (1) the formation of a small meniscus below the droplet at short times $t \lesssim 0.5$~s and (2) the slow diffusion of molecules diffusion across the oil-water interface, which gradually reduces the droplet volume over times $t\gtrsim 10^4$~s.
For a spherical cap of constant volume $V=V_S$, the contact angle $\theta(t)$ is prescribed by the contact radius $R(t)$ through the geometric relation 
$V=R^3 f_V(\theta)$, where $f_V(\theta)=\pi\left(\textstyle{\frac{2}{3}}-\textstyle{\frac{3}{4}}\cos\theta+\textstyle{\frac{1}{12}}\cos 3\theta\right)/\sin^3\theta$.
Hence, from the equilibrium radius $R_E=\sqrt[3]{V/f_V(\theta_E)}$ and height $h_E=R_E(1-\cos\theta_E)/\sin\theta_E$ observed for times $t\simeq 10^3$ to $10^4$ s (cf. Fig.~\ref{fig:1}d) we estimate apparent equilibrium contact angles $\theta_E=75\pm 4^\circ$ (see Fig.~\ref{fig:1}f), which agrees closely with values determined by Wilhelmy plate measurements (see Methods section).
Although chemical equilibrium is attained for $t > 10^4$ when the studied droplets diffuse into the ambient phase  (cf. Fig.~\ref{fig:1}c--d), the equilibrium contact angle $\theta_E$ was steadily observed for over 24 hours in Wilhelmy plate measurements. 
Our theoretical analysis will focus on the late spreading regime leading to the equilibrium configuration prescribed by the (size-independent) contact angle $\theta_E$, which corresponds to a state of mechanical equilibrium for a droplet of constant volume. 

\vskip 12 pt
\noindent {\bf \large From power-law spreading to physical ageing}\\
As reported in Fig.~\ref{fig:1}d, power-law behaviors with an expected range of exponents ($\alpha=$~0.1 to 1) are observed during the initial $\sim$0.1 seconds of the spreading process, after which there is a crossover to an unexpectedly slow regime that persists for around $10^3$ seconds until attaining the mechanical equilibrium condition prescribed by $\theta_E$.
The evolution of the droplet contact radius in the late spreading regime for times $t\simeq$~0.1--1000~s is nearly logarithmic in time and resembles the physical ageing phenomenon first reported by Kaz {\it et al.}\cite{kaz2012} for colloidal particle adsorption.
To rationalize the observed regime crossover and unexpectedly slow spreading rates we adapt the theory developed by Colosqui {\it et al.}\cite{colosqui2013} for the near-equilibrium adsorption of colloidal particles to the studied case of near-equilibrium droplet spreading. 
The spreading of a hemispherical droplet of constant volume can be described in terms of a single variable (e.g., the contact radius, contact angle, or droplet height).
While we choose the contact radius $R(t)$ to describe the spreading evolution, it is convenient to formulate certain analytical expressions in terms of the contact area $A(t)=\pi R^2$. 
Assuming a closed isothermal system in thermal and chemical equilibrium during the spreading of a hemispherical droplet on a perfectly flat and chemically homogeneous surface, the Helmholtz free energy is given by 
\begin{equation}
{\cal F}_0= \left(\frac{2}{1+\cos\theta}- \cos\theta_E\right) \gamma A.
\label{eq:fenergy0}
\end{equation}
From equation (\ref{eq:fenergy0}) one can readily obtain the effective force $F=-(d{\cal F}_0/dA)\times (dA/dR)=-2\pi R \gamma (\cos\theta-\cos\theta_E)$ driving the spreading of perfectly hemispherical droplets on ideally smooth surfaces.

Following ideas from prior work\cite{colosqui2013,rahmani2016,prf2016,keal2017}, we consider that three-dimensional topographic features with a characteristic base area $A_d\sim {\cal O}$(1 nm$^2$) induce spatial energy fluctuations $\Delta {\cal F}\sim \gamma A_d$ that are neglected in the free energy expression (equation (\ref{eq:fenergy0})) for a perfectly flat surface.
Accordingly, the free energy for a droplet spreading on a surface that is densely populated with nanoscale topographic features with characteristic base area $A_d$ can be approximately modeled as
\begin{equation}
{\cal F}= {\cal F}_0
+\frac{1}{2}\Delta{\cal F} 
\sin\left( \frac{2\pi A}{A_d}+\phi\right),
\label{eq:fenergy}
\end{equation}
where $\Delta{\cal F}$ is the characteristic magnitude of energy fluctuations induced by the nanoscale topography, and the phase $\phi \in [0,2\pi)$ can be arbitrarily chosen given that $A_d\ll A$.   
For overdamped Markovian dynamics (i.e., neglecting inertia and memory effects) the evolution of the contact area can be described by a Langevin equation 
\begin{equation}
\xi_A \frac{dA}{dt}=-\frac{d{\cal F}}{dA}+\sqrt{2 k_B T \xi_A} \eta(t),
\label{eq:langevin}
\end{equation}
where $\xi_A(A)$ is the damping coefficient determining the dissipative force, $k_B$ is the Boltzmann constant, and the random function $\eta$ is zero-mean unit-variance Gaussian noise. 
The random term in equation (\ref{eq:langevin}) is a mathematical ansatz designed to satisfy the fluctuation-dissipation theorem\cite{kubo1966} and it is included to model thermal fluctuations of the contact area $A$.
The energy dissipation rate is $dE/dt=-\xi_A (dA/dt)^2=-\xi (dR/dt)^2$ and thus $\xi_A=\xi/(2\pi R)^2$ can be determined from the damping coefficient $\xi=\xi(R)$ predicted by available models for contact line dynamics.\cite{voinov1976,cox1986,blake1969}

As the system approaches the expected equilibrium at $A_E=\pi R_E^2$ and $d{\cal F}_0/dA|_{A_E}\to 0$, the contact area evolution described by equation (\ref{eq:langevin}) becomes a sequence of random thermally activated transitions between neighboring metastable states corresponding to local minima in equation (\ref{eq:fenergy}).
Sufficiently close to equilibrium the noise-averaged evolution of the contact area governed by equation (\ref{eq:langevin}) can be described by the rate equation\cite{colosqui2013} 
\begin{equation}
\frac{dA}{dt}=A_d \left(\Gamma_+-\Gamma_-\right),
\label{eq:rate}
\end{equation}
where according to Kramers theory\cite{kramers1940,hanggi1986} the forward/backward (+/-) transition rates are
\begin{eqnarray}
\label{eq:kramers}
\Gamma_\pm&=&\frac{\sqrt{(\pi/A_d)^4 (\Delta{\cal F}/2)^2-K^2}}{2\pi\xi_A}\\ \nonumber
&\times&\exp\left(-\frac{\Delta{\cal F}+K A_d^2/8}{k_B T} \right)
\exp\left[\pm \frac{K A_d (A_E-A)}{2 k_B T} \right].
\end{eqnarray}
Here, the parameter $K=d^2{\cal F}_0/dA^2|_{A_E}$ is the curvature at equilibrium of the free energy (equation (\ref{eq:fenergy0})) for a perfectly flat surface.
Analytical integration of equation (\ref{eq:rate}) for the case that $A_E\gg 2 k_B T/K A_d$ (see Supplementary Information) gives the implicit relation
\begin{equation}
\frac{t}{T_K} =
-\frac{R_E}{R} \log\left[\tanh \left( \frac{R_E^2-R^2}{R_K^2}\right)\right]
\label{eq:area}
\end{equation}
between the contact radius and time.
In equation (\ref{eq:area}) we have introduced the characteristic `kinetic'' time
\begin{eqnarray}
\label{eq:Tk}
T_K&=&\frac{A_d}{2}
\frac{\xi_E R_K^2}{\sqrt{\pi^4(\Delta{\cal F}/2)^2-K^2 (A_d/2\pi)^4}}\\ \nonumber
&\times& \exp\left(\frac{\Delta{\cal F}+K A_d^2/8}{k_B T}\right),
\end{eqnarray}
%
%
the ``kinetic'' length $R_K=\sqrt{2 k_B T/K \pi A_d}$, and the equilibrium damping coefficient $\xi_E=\xi(R_E)$.
It is worth noticing that equation (\ref{eq:area}) predicts a slow logarithmic evolution of the contact area 
\begin{equation}
R^2=R_E^2+R_K^2 \log(t/2T_K)
\label{eq:logarithmic}
\end{equation} 
in the near-equilibrium spreading regime for which $|R-R_E|\ll R_E$. 
The predicted logarithmic droplet evolution near equilibrium over long times $T_K$ is analogous to the physical ageing phenomenon reported for microparticles at a water-oil interface.

A slow kinetic regime is described by the implicit expression in equation (\ref{eq:area}) or the explicit logarithmic expression in equation (\ref{eq:logarithmic}), which are derived from the rate equation (equation (\ref{eq:rate})) for thermally activated transitions between metastable states.
Such metastable states correspond to local minima in the free energy profile ${\cal F(R)}$ given by equation (\ref{eq:fenergy}), which can only exist when the droplet is sufficiently close to equilibrium and $K|R^2-R_E^2|\le\Delta{\cal F}/A_d$. 
Accordingly, the kinetic spreading regime governed by equation (\ref{eq:area}) should be only be observed for contact radii $R>R_C$ larger than the crossover radius
\begin{equation}
R_C=\sqrt{\left|R_E^2-\alpha \frac{\Delta{\cal F}}{K A_d}\right|},
\label{eq:crossover}
\end{equation}
where $\alpha\simeq 0.5$, based on numerical analysis and experimental observations for different systems.\cite{colosqui2013,rahmani2016,keal2017,prf2016} 

\begin{table}
\caption{Model parameters for analytical fits}
\begin{center}
\vskip -7pt
\begin{tabular*}{0.45\textwidth}{@{\extracolsep{\fill}}cccccc}
\hline
$V$ [$\mu$L] & 0.24 & 0.34 & 0.37 & 0.41 & 0.63\\
\hline
$A_d$ [nm$^2$] &	4.2 &	4.2 &	4.2 & 4.2 & 4.2\\
$\Delta{\cal F}/k_B T$ & 11.5 & 16.8 & 15.9 & 15.9 & 14.2\\
$\chi=\xi/\mu_o 2\pi R$ & 160 & 175 & 175 & 150 & 111 \\
\end{tabular*}
\end{center}
\vskip -20 pt
\end{table}

\vskip 12 pt
\noindent {\bf \large Experimental results vs. model predictions}\\
\vskip -12 pt
\noindent Experimental results and analytical predictions for microliter droplets of different volumes ($V=$~0.24 to 0.63 $\mu$L) are reported in Fig.~\ref{fig:2}.
The initial spreading dynamics for $R<R_C$ can be described by power-law scalings $R\propto t^\alpha$ with exponents $\alpha\simeq$~$2/3$ to 1 (see also Fig.~\ref{fig:1}d), which can be accounted for by using damping coefficients estimated by MKT (see Supplementary Information) in equation (\ref{eq:langevin}).
For the studied near neutral wetting conditions with a viscous ambient phase, Tanner's law $\alpha\simeq 0.1$ does not agree with the experimental observations (see Fig.~\ref{fig:2}a-c). 
The time evolution of the contact radius (Fig.~\ref{fig:2}a-c) shows a crossover from power-law behavior to a nearly logarithmic regime described by the implicit relation in equation (\ref{eq:area}) and the approximate explicit expression in equation (\ref{eq:logarithmic}).
The parameters employed to produce the analytical fits are reported in Table 1.
A base defect area $A_d=4.2$ nm and energy barriers $\Delta {\cal F}= 14.1\pm 2.7 k_BT$ account for the spreading rates in the kinetic regime and the crossover radius $R_C$ for which the regime crossover is observed (cf. Fig.~\ref{fig:2}).
As expected, the crossover radius $R_C$ can be estimated in all cases by equation (\ref{eq:crossover}) for $\alpha=0.55$.
While the regime crossover is attributed to the emergence of local minima in the free energy profile, and thus it is independent from dissipative effects, the spreading rates in the late kinetic regime are influenced by dissipative forces, which are determined by the effective damping coefficient $\xi(R)$.  
We estimated damping coefficients $\xi(R)\simeq -2\pi\gamma R (cos\theta-\cos\theta_E) /(dR/dt)$ from observed contact radius displacement rate $dR/dt$ and apparent contact angle $\theta(R)$ in the initial spreading dynamics where metastable states are not present.
The damping coefficients thus estimated from our experimental observations (see Supplementary Information) are in close agreement with MKT predictions,\cite{blake1969,blake2006} according to which 
$\xi(R)=\chi \mu_o 2\pi R$ where the friction factor $\chi=(\nu/\lambda^3)\exp(\gamma \lambda^2 (1+\cos\theta_E)/k_BT)$ is prescribed by the characteristic molecular volume $\nu$ and adsorption site size $\lambda$.
Friction factors $\chi=$~111 to 175 accounting for our experimental observations (see Table 1) can be estimated by using $\nu=1.27\times 10^{-28}$~m$^3$ for the studied PDMS oil and assuming molecular adsorption site sizes $\lambda=$~0.63 to 0.67 nm, which are close to the values reported in previous studies using MKT.\cite{blake2006,de1997contact,ramiasa2014}  
Employing MKT and the friction factors estimated from the initial spreading dynamics we determine the equilibrium damping coefficients $\xi_E=\chi\mu_o 2\pi R_E$ employed in equation (\ref{eq:Tk}) for the kinetic relaxation time in late regime.

\begin{figure}[h!]
\center
\includegraphics[angle=0,width=1.\linewidth]{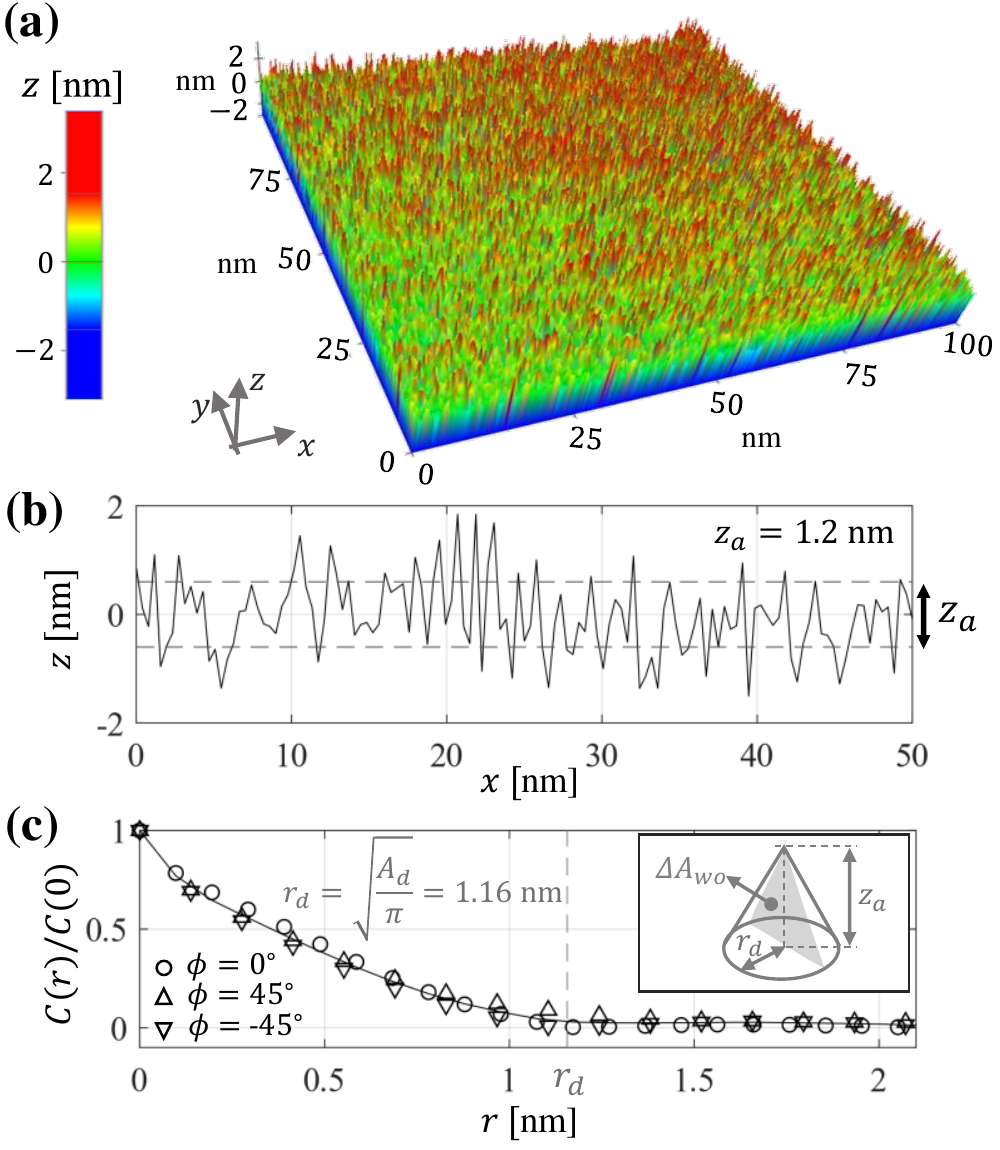}
\vskip -5pt
\caption{Nanoscale surface topography and model parameters.
(a) 3D topographic image (512$\times$512 pixels) of a borosilicate glass employed for spreading experiments obtained via NC-AFM.
(b) 1D local height profile and average height $z_a=2\int_0^\ell |z| dx/\ell$ = 1.2 nm.
(c) Height autocorrelation function for different directions $\phi={\rm atan}(y/x)=$ 0,$\pm$ 45 deg. Vertical dashed line indicates the correlation length $r_d=\sqrt{A_d/\pi}$ corresponding to averaged topographic ``defects'' with base area $A_d=4.2$~nm$^2$.
Inset illustrates a modeled conical ``defect'' with base area $A_d=\pi r_d^2$, height $z_a$, and cross-sectional area $\Delta A_{wo}$.
}
\label{fig:3}
\end{figure}

As reported in Fig.~\ref{fig:2}d, the evolution of the contact radius of droplets of different volume after the crossover to the late kinetic regime can be collapsed onto a single curve where the evolution time is normalized by the kinetic time $T_K$ determined by equation (\ref{eq:Tk}).
The proper combination of base ``defect'' area and energy barrier is required to account for the regime crossover position predicted by equation (\ref{eq:crossover}) and the relaxation rates in the near-equilibrium regime.
The fact that a single defect area $A_d=4.2$~nm$^2$ and narrow range of energy barriers $\Delta {\cal F}\simeq 14.1 \pm 2.7 k_B T$ (see Table 1) accounts for experimental results for droplets of different volume suggests that the nanoscale topography of the substrate may strongly influence the observed near-equilibrium spreading.
We therefore seek to determine whether the base area and energy barriers used for analytical fits are indeed related to roughness parameters such as the radial correlation length and the average height.\cite{gadelmawla2002}
Atomic Force Microscopy (AFM) in non-contact mode (see Methods section) was employed to produce (512 $\times$ 512 pixels) topographic images of 100 $\times$ 100 nm sections of the borosilicate glass substrates employed in the experiments (see Fig.~\ref{fig:3}a).
The analysis of topographic profiles obtained via AFM reveals the presence of random nanoscale roughness having a nearly isotropic and Gaussian height distribution with average amplitude $z_a=2\int_0^\ell |z| dx/\ell=0.6$ nm (see Fig.~\ref{fig:3}b), standard deviation $\sigma\simeq 0.75$~nm, small positive skewness $\zeta\simeq 0.3$, and excess kurtosis $\kappa\simeq 0.1$.
The radial autocorrelation function 
$C(r)=\lim_{\ell\to\infty} \int_0^\ell z(\tau) z(\tau+r) d\tau/\ell$ 
computed from AFM data (Fig.~\ref{fig:3}c) shows a similar nearly exponential decay in different directions $\phi={\rm{atan}}(y/x)$. 
Notably, the decay of the height autocorrelation function is characterized by a correlation length 
$r_d\simeq 1.16$~nm, which indicates that topographic features approximately have the average base area $A_d = \pi r_d^2 \simeq 4.2$~nm employed for analytical fits (see Table 1).
Furthermore, by modeling topographic ``defects'' as cones with a base radius $r_d$ and height 
$ z_a$ (see inset in Fig.~\ref{fig:3}c) one can estimate a characteristic energy barrier $\Delta{\cal F}\simeq \gamma \Delta A_{wo}=14.6~k_B T$ associated to the small area variation $\Delta A_{wo}=r_d\times z_a$ that occurs when the oil-water interface moves over a single ``defect''.
Hence, we find that the characteristic energy barrier estimated from the average geometric dimensions of 3D topographic features agrees closely with the mean energy barrier employed in the analytical fits reported in Fig.~\ref{fig:2} (see Table 1).

\vskip 12 pt
\noindent{\bf \large Summary and outlook}\\
In summary, during the early dynamics of droplet spreading extending for about 0.1 s we observe power-law behaviors governed by capillary forces and effective damping forces that can be rationalized by MKT, as reported in previous studies.\cite{blake2006,de1997contact,ramiasa2014}
The damping coefficient predicted by MKT solely considers the viscosity of the ambient phase, which is aobut 100 times larger than the droplet viscosity in our experiments.
The late near-equilibrium behavior, hoewever, does not follow a power law and exhibits a nearly logarithmic-in-time evolution $R(t)^2 \propto \log t$ with characteristic times to reach equilibrium on the order of thousands of seconds. 
The observed late spreading behavior resembles the physical ageing phenomenon previously reported for colloidal microparticles at liquid-liquid interfaces.\cite{kaz2012,rahmani2016,keal2017} 
The crossover between fast and slow spreading regimes is analytically estimated by considering that close to equilibrium the one-dimensional free energy profile becomes densely populated with metastable states having a characteristic period and energy barrier prescribed by the nanoscale topography of the substrate. 
The spreading rates in the near-equilibrium regime are estimated by using Kramers theory\cite{kramers1940,hanggi1986,colosqui2013} for thermally activated escape from metastable states.
Notably, the spreading model proposed in this work yields quantitative agreement with experimental observations when employing as input parameters the average base area  $A_d =$ 4.2 nm$^2$ and average height of $z_a \simeq 1.2$~nm of nanoscale defects observed by AFM topographic analysis.

The findings in this work indicate that physical topographic features induce the slow thermally activated spreading that is observed when the studied droplets are near mechanical equilibrium. 
The proposed model based on a one-dimensional energy profile with a single-mode perturbation of amplitude $\Delta{\cal F}\simeq \gamma z_a \sqrt{A_d/\pi}$ and period $A_d$ is able to quantitatively predict both the crossover to the slow kinetic regime and the spreading rate in the final approach to equilibrium.  
An important implication of this work is that according to equation (\ref{eq:crossover}) certain combinations of defect height and base area could induce the crossover to the slow kinetic regime at very early stages in the spreading process, which would effectively hinder the spreading and adhesion of droplets with a specific range of volumes on surfaces with different wettability.   
Future experimental and computational work employing different liquid pairs with varying viscosity ratios and nanostructured surfaces with well-characterized topographic features can be readily designed to verify this prediction.

\vskip 10pt
\noindent{\bf \large Methods}\\
\noindent {\bf Spreading experiments.}
The working fluids are DI water (Sigma Aldrich 38796) and 100-cSt silicone oil (Sigma Aldrich 378364).
In order to gently deposit single millimeter-sized droplets a small water volume (0.3 to 0.5 ~$\mu$L) is dispensed by a syringe pump to produce a spherical drop suspended at the capillary tip (cf. Fig.~\ref{fig:1}b). 
After a suspended drop is formed, the capillary is manually pulled upward toward the oil-air interface (see rightmost panels in Fig.~1b) in order to induce the detachment of the drop from the capillary and its subsequent deposition on the borosilicate glass slide where the spontaneous spreading process takes place.
The volume of the deposited droplet differs from the injected volume because a small uncontrollable volume of the liquid filament inside the capillary becomes part of the droplet after detachment.
The deposition and spreading process were recorded using digital imaging from a lateral wall of the immersion cell with a high-speed camera (AOS Technologies AG QPRI) up to 1000 fps and time-lapse photography at 1/500 fps.
Digital processing of the acquired image sequences was performed using the public domain software ImageJ.\cite{schneider2012nih}
The spreading of droplets of various differetn volumes was recorded during 130 seconds using imaging rates between 32 and 50 fps, which allowed for efficiently resolving the crossover from initial (fast) to late (slow) spreading regimes.
A few experiments were recorded at 500 to 1000 fps during the first 10 seconds, after which time lapse photography at 500 second intervals was employed for up to 3 days to capture the entire evolution to mechanical equilibrium.
The employed borosilicate glass slides (McMaster-Carr) are cleaned with DI water, heated at 400$^\circ$C for 1 hr and allowed to return to room temperature inside an oven, after which dry air is blown to remove dust particles before placing them in the oil bath.

\noindent {\bf Contact angle characterization.}
Contact angles were determined via the Wilhelmy plate (force-displacement) method using a force tensiometer (Sigma 700 by Biolin Scientific). Force measurements for borosilicate slides and working fluids employed in spreading experiments were performed continuously for 30 hours using very low displacement speeds $V =$ 0.01 mm/min. After the relaxation of the water-oil meniscus, advancing and receding contact angles remained steadily within the range 74.5$\pm$ 3$^\circ$ over a 24-hour period.   

\noindent {\bf AFM topographic imaging.} Topographic images of borosilicate glass slides employed in the spreading experiments were performed at the Center for Functional Nanomaterials in Brookhaven National Laboratory. Measurements were performed using a Park NX-20 AFM in Non-Contact (NC) mode and cantilever probes PPP-NCHR by Park Systems. NC-AFM images were obtained for square sections of 100$\times$100 nm and 50$\times$50 with resolutions varying from 256$\times$256 pixels to 1024$\times$1024 pixels, which produce similar statistical properties.

\noindent {\bf Data availability.} Data reported in this work will be made available upon request to the corresponding author [C.E.C.]. 

\vskip 5pt
\noindent{\bf \large References}
\vskip -10pt

%

%
%

\vskip 7pt
\noindent{\bf \large Acknowledgments}\\
%
\noindent This work has been supported by the National Science Foundation CBET-1605809 and has used resources of the Center for Functional Nanomaterials (CFN), which is a U.S. DOE Office of Science Facility, at Brookhaven National Laboratory under Contract No. DE-SC0012704.
N.D. was partially supported by Fellowship from Joint Photon Sciences Institute at Stony Brook University.
We thank Xiao Tong and Dario Stacchiola at CFN for technical advice on performing high-resolution AFM imaging.

\vskip 7pt
\noindent{\bf \large Author contributions}\\
%
\noindent B.M.J. and T.C. designed the experimental apparatus for spreading experiments, conducted the spreading experiments and produced data sets employed in the analysis. 
D.N. performed Wilhelmy plate measurements and AFM topgraphic images and assisted with analysis and visualization of experimental data. 
C.E.C. led the development of the analytical model and interpretation of experimental
data.  
All authors contributed to the preparation of the manuscript.
%

\vskip 7pt
\noindent{\bf \large Competing financial interests}\\
\noindent The authors declare no competing financial interests.

\vskip 7pt
\noindent{\bf \large Additional information}\\
\noindent {\bf Supplementary information} is available for this manuscript.\\
\noindent {\bf Correspondence and requests for materials} should be addressed to C.E.C.

\end{document}